\documentclass[referee,useAMS,usenatbib,usegraphicx]{mn2e}

\usepackage{amssymb}
\usepackage{graphicx}

\title[Distance estimation of Bok globules]{Distance estimation of some selected small Bok globules}
\author[A. Das, H. S. Das and A. S. Devi]
{A. Das$^{1}$, H. S. Das$^{1}$\thanks{E-mail: hsdas@iucaa.ernet.in (HSD), corresponding author}
, A. S. Devi$^{1}$\\
\\
$^{1}$Department of Physics, Assam University, Silchar 788011, India\\
}

\begin{document}

\date{Accepted 2015 June 5. Received 2015 June 2; in original form 2015 March 29}

\pagerange{\pageref{firstpage}--\pageref{lastpage}} \pubyear{2015}

\maketitle

\label{firstpage}

\begin{abstract}
The distance determination of small Bok globules is often difficult because of its small size and opaqueness. In this work, we determine the distances to six small Bok globules CB 17, CB 24, CB 188, CB 224, CB 230 and CB 240 using near infra red photometry (2MASS $JHK_S$ colors). The distances to these clouds are estimated to be $478 \pm 88$, $293 \pm 54$, $262 \pm 49$, $378 \pm 70$, $293 \pm 54$ and $429 \pm 79$ parsec respectively.

\end{abstract}


\section{Introduction}
The distance determination to interstellar clouds is important to estimate the sizes, masses and densities of the cloud \citep{cyh} and to obtain luminosities of the protostars embedded in the clouds \citep{yc}. The distances are sometimes difficult to estimate with reasonable accuracy. Generally, two methods are adopted to determine the distance \citep{wpf}. One is to determine the distances to selected individual stars using spectrophotometric technique which are known to be at the same distance as the cloud. The second method is to study the dependency of interstellar reddening on distances for a large sample of field stars distributed along the line of sight. In this technique, the sudden rise of extinction of a field star at a particular distance is assumed to be the distance of the cloud. The second approach can work well for isolated clouds at intermediate galactic latitude.

A number of techniques adopted by various investigators in past to estimate the distances to clouds which are either associated with large globules or are relatively isolated \citep{gs,r1,bc,s2,rg,kp,cfk,pc,f1,mmb,af,ltm,mlb,emp,dlm,bd}. Using Near Infrared (NIR) photometric method, \cite{mlb} determined the distances to four molecular clouds L1517, Chamaeleon I, Lupus 3 and NGC 7023 which are found to be in good agreement with the most accurate distances available for them in literature. Later, \cite{mld}  and \cite{emp} used the same technique to determine the distances to couple of clouds.  Recently, \cite{bd} adopted Maheswar's technique to estimate the distance to the small Bok globule CB 4 which is given by $459 \pm 85$ parsec (pc). This method finds the nearest cloud layer along the line of sight.

In the present work, the distances to some small isolated CB clouds have been determined using the technique developed by \cite{mlb}.

\section{Distance determination of dark clouds}
Bok globules are small, opaque, isolated simply structured molecular clouds that often contain only one or two star forming cores. Several catalogues of dark clouds and globules were published during the last couple of decades  \citep{sl,fs,hms,cb,pfb}. Among them catalogue published by \cite{cb} (CB catalogue) is most homogeneous and complete collection of northern globules. The distances to the CB clouds are often difficult to determine because they are small and too opaque to apply star counts or photometric methods as distance estimators. It is noticed that the absence of foreground stars towards many globules with angular diameters of $2^{\prime} - 8^{\prime}$  suggests that these globules can not be further away then about 500 pc \citep{bc}. The globules which are located much further away can not be distinguished on optical images against the stellar background and so most of the Bok globules are assumed to be situated within 1 kpc around the Sun in the local spiral arm \citep{lh}. Generally, the distance of the globules are estimated with a method which associates the globules with larger molecular cloud complexes. The derived distances range from 140 pc to 1.5 kpc \citep{ln}.

In the present work, we determine the distances to Bok globules using the NIR photometric method developed by \cite{mlb}. This technique is one of the powerful technique which uses vast $J H K_s$ photometric data from 2MASS catalogue and can give distances to globules $\sim 500$ parsec with a precision of $\sim 18\%$. The detailed description of the method and its validity are discussed in \cite{mlb}. We selected the CB clouds from the catalogue prepared by \cite{cb} keeping in mind that the distance of the clouds is $\sim 500$ pc and their distance is already known. We compiled the distances of CB clouds from several sources (Peterson \& Clemens 1998; Launhardt et al. 2010 and references therein). The selected twelve clouds are CB 17, CB 24, CB 26, CB 68, CB 130, CB 188, CB 199, CB 224, CB 230,  CB 240, CB 244 and CB 246.

We collected the $J$, $H$ and $K_s$ magnitudes of stars from the 2MASS all Sky Catalog of Point Sources \citep{cs} which satisfies the following criteria:

\begin{enumerate}
  \item photometric uncertainty $\sigma_J$, $\sigma_H$, $\sigma_{K_S} \le 0.035$  in all three filters,

  \item  a photometric quality flag of ``AAA" in all three filters which corresponds to signal-to-noise ratio (SNR) $>$ 10, and

  \item $(J - K_s) \leq 0.75$ to eliminate M-type stars from the analysis.
\end{enumerate}

In our work, we considered $10^\prime \times 10^\prime$ field-of-view (FOV) for individual cloud to search for $J$, $H$ and $K_s$ magnitudes from 2MASS catalogue. The selected field stars in the cloud must satisfy the above conditions and are presented in Table-1.  It can be seen from Table-1 that the clouds CB 130  and CB 244 contain only 2 and 3 stars respectively, so we discarded them from our analysis.

In Maheswar's technique, some of the stars classified as main sequence could actually be giants which may lead to false distance to cloud. This problem could be solved if one divides the cloud into four small sub-fields (each field of $5^\prime \times 5^\prime$ in area is taken), because the rise in the extinction due to the presence of a cloud should occur almost at the same distance in all four fields. In Fig. 1, the field of a cloud with area $10^\prime \times 10^\prime$  has been divided into the four fields each with FOV $5^\prime \times 5^\prime$.


\begin{figure}
\begin{center}
\includegraphics[width=100mm]{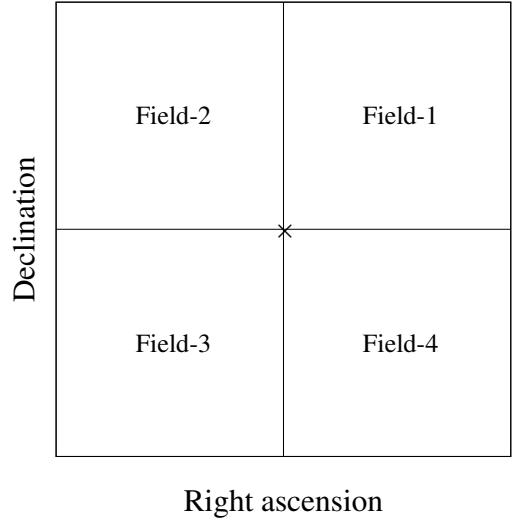}
 \caption{The field of a cloud with $10^\prime \times 10^\prime$ area has been divided into the four fields each with FOV $5^\prime \times 5^\prime$. The symbol `$\times$' denotes the central coordinates of the cloud.}
 \end{center}
\end{figure}


\begin{table}
\begin{center}
\caption{Coordinates and distances of the selected Clemens \& Barvainis (CB) globules. The seventh column represents number of stars ($N$) within $10^\prime \times 10^\prime$ field area which satisfy the criteria $(i), (ii)$ and $(iii)$ discussed in \emph{Section-2}.   }
\vspace{1cm}
\begin{tabular}{|c|c|c|c|c|c|c|c|}
  \hline
  S/N & Name & Other names &   RA     & DEC       & Distance (ref) & $N$ & Remarks  \\
      &      &            & (2000)   & (2000)    &  (in parsec)        &  &       \\
      \hline
  1    &  CB 17    & L 1389        & 04:04:38.0	& 56:56:11     & $250 \pm 50$ (1)      &   35   &  ... \\
  2    &  CB 24    & ...           & 04:58:29.7	& 52:15:41     & 360 (2)               &   23   &  The reported distance \\
       &           &               &            &              &                    &        &  corresponds to maximum  \\
       &           &               &            &              &                    &        &  distance of the cloud. \\
  3    &  CB 26    & L 1439        & 05:00:09.3	& 52:05:00     & 140 (3)               &   12   &  ... \\
  4    &  CB 68    & L 146         & 16:57:16.3	& --16:09:22   & 160 (4)               &   10   &  ... \\
  5    &  CB 130    &L 507         & 18:16:14.7	& --02:32:47   & 200 (4)               &   02   &  ... \\
  6    &  CB 188    & ...          & 19:20:16.9	& 11:36:12     & 300 (4)            &   12   &  ...   \\
  7    &  CB 199    & B 335, L 663 & 19:36:58.7	& 07:34:32     & 100 (5)                &   35   &  ...   \\
  8    &  CB 224    & L 1100       & 20:36:17.2	& 63:53:15     & 400 (6)               &   15   &  ...   \\
  9    &  CB 230    & L 1177       & 21:17:40.9	& 68:18:23     & 400 (7, 8)              &   10   &  ...   \\
  10    & CB 240     & L 1192      & 22:33:48.4	& 58:33:30     & 500 (4)               &   55   &  The reported distance \\
        &           &              &            &              &                    &        &  is an estimate on the upper\\
        &           &              &            &              &                    &        &  limit of the distance \\
        &           &              &            &              &                    &        &  which is rather uncertain.\\
  11    & CB 244     & L 1262      & 23:25:48.8	& 74:17:37     & 200 (9)               &   03   &  ...   \\
  12    & CB 246     & L 1253      & 23:56:43.6	& 58:34:29     & 140 (4)               &   20   &  ...   \\
  \hline
\end{tabular}
\end{center}
\begin{center}
\end{center}
\hspace*{5cm} \underline{References}:

\hspace*{5cm} 1. \cite{ln}

\hspace*{5cm} 2. \cite{pc}

\hspace*{5cm} 3. \cite{ls}

\hspace*{5cm} 4. \cite{lh}

\hspace*{5cm} 5. \cite{oo}

\hspace*{5cm} 6. \cite{l1}

\hspace*{5cm} 7. \cite{wlh}

\hspace*{5cm} 8. \cite{kzs}

\hspace*{5cm} 9. \cite{hl}
\end{table}

In this technique, a set of dereddened colours for each star is produced from their observed colours by using the trial values of $A_V$  in the range $0-10$ mag and the reddening law of \cite{rl}. The calculated set of dereddened color indices are then compared with the intrinsic color indices of normal main-sequence stars. The best fit values of the dereddened colors to the intrinsic colors giving a minimum values of $\chi^2$ ($\le 0.1$) can yield the corresponding spectral type and $A_V$ for the star.

The photometric distances of the stars are then estimated using the equation:

\begin{equation}
d\; \textrm{(in pc)} = 10^{(K_S-M_{K_S}+5-A_{K_S})/5}
\end{equation}
where $K_S$, $M_{K_S}$ and $A_{K_S}$ ($= 0.112\times A_V$) are apparent magnitude, absolute magnitude and extinction, respectively.

The uncertainty in $A_V$ is given by
\begin{equation}
\sigma(A_V) = \sqrt{4.7^2.\sigma^2_{JH} + 7.9^2.\sigma^2_{HK_S} + 2\times37\times cov(JH,HK_S)}
\end{equation}
where $\sigma^2_{JH} = \sigma^2_{J}+\sigma^2_{H}$, $\sigma^2_{HK_S} = \sigma^2_{H}+\sigma^2_{K_S}$ and $cov(JH, HK_S) = r_s \times \sigma_{JH} \sigma_{HK_S}$. The Spearman rank-order correlation coefficient ($r_s$) is calculated from uncertainties in $(J-H)$ and $(H-K_S)$ colors which shows a strong correlation between them.

Also, the uncertainty in distance is estimated using the expression,
\begin{equation}
    \sigma_d = \sqrt{(\sigma^2_{K_S} + \sigma^2_{M_{K_S}} + \sigma^2_{A_{K_S}})\times (d/2.17)^2}
\end{equation}
where $\sigma_{K_S}$ is the uncertainty in $K_S$ band, $\sigma_{M_{K_S}}$ is the uncertainty in the estimation of the absolute magnitude and $\sigma_{A_{K_S}}$ is the uncertainty in the $A_{K_S}$ estimated by the method. The uncertainty in the $\sigma_{M_{K_S}}$ is assumed to be 0.4 while calculating $\sigma_d$ in the distances for all the stars.

The distance versus extinction plot will be generated for selected clouds. The first star which shows sudden rise in extinction in the plot will be considered as the distance to the cloud. We also assume that the extinction of distance indicator star should be more than 0.5.

We discarded four clouds CB 26, CB 68,  CB 199 and CB 246 from our study. The distances reported by \cite{ln} were 140 pc, 160 pc, 100 pc and 140 pc respectively. We determined the distances to all the field stars of these clouds, but unfortunately no field stars have been found close to the distance reported in literature for above four clouds.

\subsection{CB 17:}
CB 17 is a small and slightly cometary-shaped globule which contains two prestellar cores and one cold \textit{IRAS} point source (IRAS 04005+5647). This is located near Perseus and associated with the Lindblad ring. \cite{lh} derived the distance of the cloud CB 17 via association in projected space and radial velocity with other Lindblad Ring clouds which have a mean distance of $\approx 300$ parsec from the Sun \citep{duc}. \cite{ln} suggested a distance of $250 \pm 50$ pc for CB 17 by combining the results obtained for possible associations of CB 17 with both the Lindblad Ring and HD 25347.

In Fig. 2, the field stars in the cloud CB17 are shown in a $10 \times 10$ arcmin$^2$ R-band DSS image of the field. The extinction and distance obtained from this technique are shown in Table-2 where F1, F2, F3 and F4 correspond to Field-1, Field-2, Field-3 and Field-4.  The extinction ($A_V$) versus distance ($d$) is then plotted for four different fields in the vicinity of CB 17 and is shown in Fig. 3. The sudden rise of extinction is noticed at $d = 478 \pm 88$ pc that corresponds to star \# 10 which is considered to be the distance of the cloud. The distance obtained from our study differs from the distance suggested by \cite{ln}.

\begin{table}
\begin{center}
\caption{Selected field stars in CB 17. The table is arranged in ascending order of right ascension. The first column represents the serial number of star, second column represents 2MASS identification number, third column represents extinction (in magnitude), fourth column represents distance (in parsec) and the last column represents the particular field.}
\vspace{1cm}
\begin{tabular}{|c|c|c|c|c|}
 \hline
	   \# & 2MASS &  $A_v$ & $d$ & Field \\	
	      &       & (mag) & (in pc)& \\	
	 \hline	
1	&	04040288+5659515	&	0.01	&	1027$\pm$ 	190	&	F1	 \\
2	&	04040426+5651366	&	1.07	&	945	$\pm$ 	174	&	F4	 \\
3	&	04040568+5656009	&	0.69	&	1174$\pm$ 	217	&	F4	 \\
4	&	04040958+5659159	&	0.22	&	528	$\pm$ 	97 &	F1	 \\
5	&	04041127+5654284	&	0.17	&	799	$\pm$ 	148	&	F4	 \\
6	&	04041335+5653068	&   0.93	&	513	$\pm$ 	95	&	F4	 \\
7	&	04041374+5651394	&	0.70    &	1724$\pm$ 	319	&	F4	 \\	
8	&	04041494+5655500	&	0.23	&	616	$\pm$ 	114	&	F4	 \\
9	&	04041503+5652357	&	0	    &	712 $\pm$ 	132	&	F4	 \\
10	&	04041572+5656463	&	1.39	&	478	$\pm$ 	88	&	F1	 \\
11	&	04041835+5700245	&	1.26	&	811	$\pm$ 	150	&	F1 	 \\
12	&	04042152+5656366	&	0	    &	493	$\pm$ 	91	&	F1	 \\
13	&	04042274+5654594	&	0.58	&	716	$\pm$ 	132	&	F4	 \\
14	&	04042710+5654599	&	0.25	&	657	$\pm$ 	122	&	F4	 \\
15	&	04042783+5700030	&	0.74	&	1560$\pm$ 	288	&	F1	 \\
16	&	04042882+5654474	&	0	    &	223	$\pm$ 	41	&	F4	 \\
17	&	04043153+5655035	&	1.17	&	670	$\pm$ 	124	&	F4 	 \\
18	&	04043547+5700289	&	1.26	&	808	$\pm$ 	149	&	F1	 \\
19	&	04043647+5700307	&	1.50	&	912	$\pm$ 	169	&	F1	 \\
20	&	04043933+5653464	&	2.16	&	1167$\pm$ 	216	&	F3	 \\
21	&	04044028+5651354	&	0.75	&	841	$\pm$ 	155	&	F3	 \\
22	&	04044259+5659199	&	0.45	&	537	$\pm$ 	99	&	F2	 \\
23	&	04044471+5700059	&	1.45	&	784	$\pm$ 	145	&	F2	 \\
24	&	04044842+5659597	&	0	    &	216	$\pm$ 	40	&	F2	 \\
25	&	04044962+5653016	&	0.45	&	755	$\pm$ 	140	&	F3	 \\
26	&	04045240+5658252	&	2.11	&	843	$\pm$ 	156	&	F2	 \\
27	&	04045625+5655124	&	0.36    &	430	$\pm$ 	79	&	F3	 \\
28	&	04045631+5659534	&	0.58	&	617	$\pm$ 	114	&	F2	 \\
29	&	04045669+5651467	&	0   	&	684	$\pm$ 	126	&	F2	 \\
30	&	04045709+5658454	&	1.31	&	612	$\pm$ 	113	&	F2	 \\
31	&	04050697+5652149	&	0.84	&	1384$\pm$ 	255	&	F3	 \\
32	&	04050993+5658392	&	1.27	&	738	$\pm$ 	136	&	F2	 \\
33	&	04051158+5652029	&	0	    &	186	$\pm$ 	34	&	F3	 \\
34	&	04051178+5656488	&	0	    &	498	$\pm$ 	92	&	F2	 \\
35	&	04051350+5659246	&	0.59	&	547	$\pm$ 	101	&	F2	 \\
  \hline

\end{tabular}
\end{center}
\end{table}

\begin{figure}
\hspace*{-2cm}
\includegraphics[width=220mm]{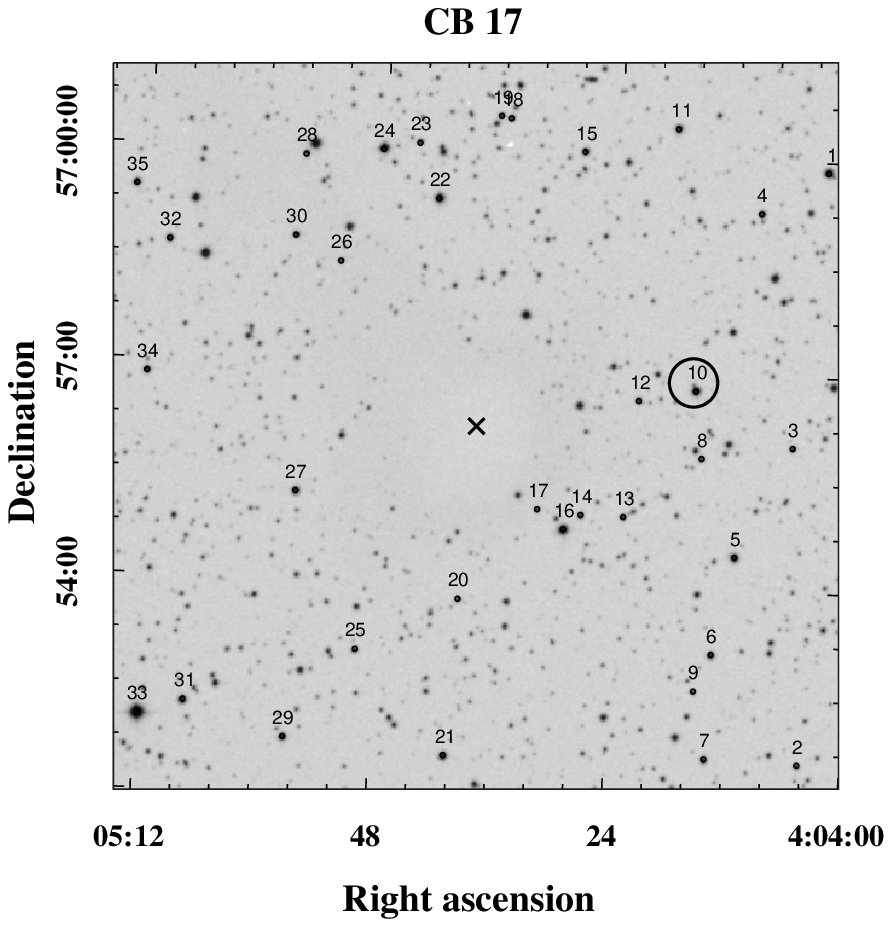}
 \caption{Selected 35 field stars in the vicinity of CB 17 shown on a $10 \times 10$ arcmin$^2$ R-band DSS image of the field. The `$\times$' symbol denotes the central coordinates of the cloud. Star \# 10 is marked by a circle which is the distance indicator of the cloud that shows sudden rise in the extinction.  }
\end{figure}

\begin{figure}
\begin{center}
\includegraphics[width=100mm]{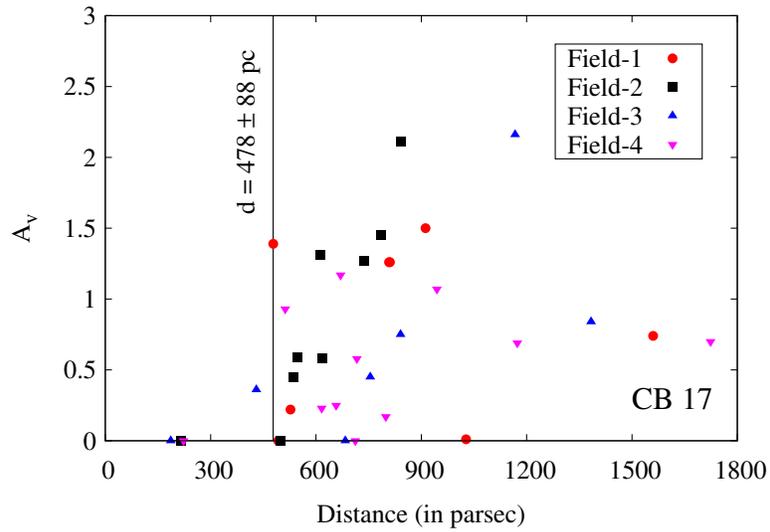}
 \caption{Extinction ($A_V$) versus distance for stars for four different fields in the vicinity of the CB 17. The vertical solid line is drawn at a distance of 478 parsec (star \# 10) where sudden rise in the $A_V$ occurs.}
 \end{center}
\end{figure}


\subsection{CB 24:}
CB 24 is a small, round shaped globule that has no associated \textit{IRAS} point source. \cite{pc} derived the distance to this cloud by identifying M dwarfs lying both in front of and behind the cloud. The maximum distance to the CB 24 was determined to be 360 pc.

In Fig. 4, the field stars in CB 24 is shown. The results obtained from this work are shown in Table-3. The $A_V$ versus $d$ plot obtained from our work  is plotted in Fig. 5. The star \# 18 is actually a foreground star with a distance of only 40 parsec. The sudden rise of extinction is obtained for star \# 10 at $ d = 293 \pm 54$ which is considered to be the distance to this cloud. Thus the maximum distance to this cloud is 347 pc which is also close to the maximum distance determined by \cite{pc}.

\begin{table}
\begin{center}
\caption{Selected field stars in CB 24.}
\vspace{1cm}
\begin{tabular}{|c|c|c|c|c|c|}
 \hline
	   \# & 2MASS &  $A_v$ & $d$ & Field \\	
	      &       & (mag) & (in pc)& \\	
\hline	
1	&	04580481+5211075	&	0.28	&	345	$\pm$	64	&	F4	 \\
2	&	04580633+5218379	&	1.47	&	1106$\pm$	204	&	F1	 \\
3	&	04581304+5217342	&	0.28	&	633	$\pm$	117	&	F1	 \\
4	&	04581536+5217027	&	2.14	&	754	$\pm$	139	&	F1	 \\
5	&	04581622+5217543	&	1.00	&	562	$\pm$	104	&	F1	 \\
6	&	04581701+5212105	&	0	    &	418	$\pm$	77	&	F4	 \\
7	&	04581767+5214592	&	0.93	&	502	$\pm$	93	&	F4	 \\
8	&	04581808+5217426	&	0.03	&	373	$\pm$	69	&	F1	 \\
9	&	04581885+5211116	&	0.55	&	351	$\pm$	65	&	F4	 \\
10	&	04582261+5216339	&	0.86	&	293	$\pm$	54	&	F1	 \\
11	&	04582327+5211503	&	0.97	&	1221$\pm$	226	&	F4	 \\
12	&	04582742+5217109	&	0	    &	175	$\pm$	32	&	F1	 \\
13	&	04583198+5218167	&	0.14	&	368	$\pm$	68	&	F2	 \\
14	&	04583296+5216340	&	1.45	&	987	$\pm$	183	&	F2	 \\
15	&	04583836+5219133	&	1.51	&	727	$\pm$	134	&	F2	 \\
16	&	04584701+5212410	&	2.07	&	1117$\pm$	206	&	F3	 \\
17	&	04584864+5220189	&	0.83	&	304	$\pm$	56	&	F2	 \\
18	&	04584964+5212165 	&	0.91	&	40	$\pm$	7	&	F3	 \\
19	&	04585065+5214172	&	1.64	&	790	$\pm$	146	&	F3	 \\
20	&	04585471+5215486	&	0.95	&	389	$\pm$	72	&	F2	 \\
21	&	04585940+5219126	&	0.09	&	191	$\pm$	35	&	F2	 \\
22	&	04585983+5220163	&	1.02	&	603	$\pm$	111	&	F2	 \\
23	&	04590112+5220100	&	1.42	&	426	$\pm$	79	&	F2	 \\
\hline
\end{tabular}
\end{center}
\end{table}

\begin{figure}
  \hspace*{-2.5cm}
\includegraphics[width=220mm]{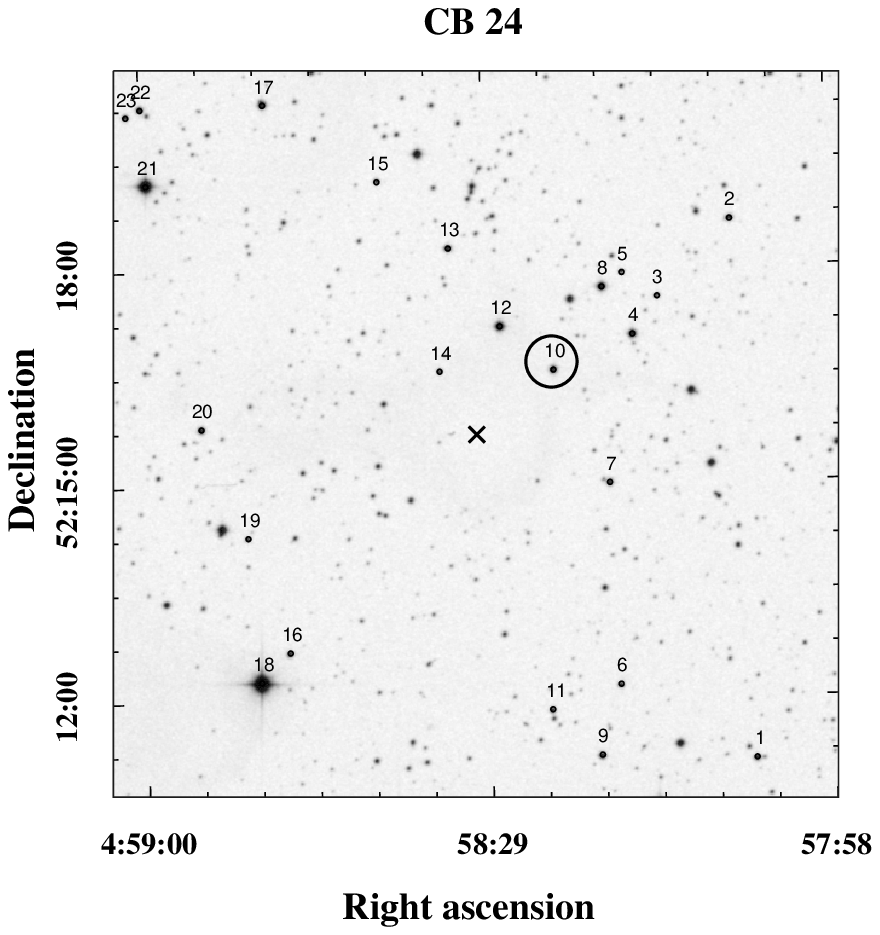}
 \caption{Selected 23 field stars in the vicinity of CB 24 shown on a $10 \times 10$ arcmin$^2$ R-band DSS image of the field. The `$\times$' symbol denotes the central coordinates of the cloud. Star \# 10 is marked by a circle which is the distance indicator of the cloud that shows sudden rise in the extinction.}
\end{figure}

\begin{figure}
\begin{center}
\includegraphics[width=100mm]{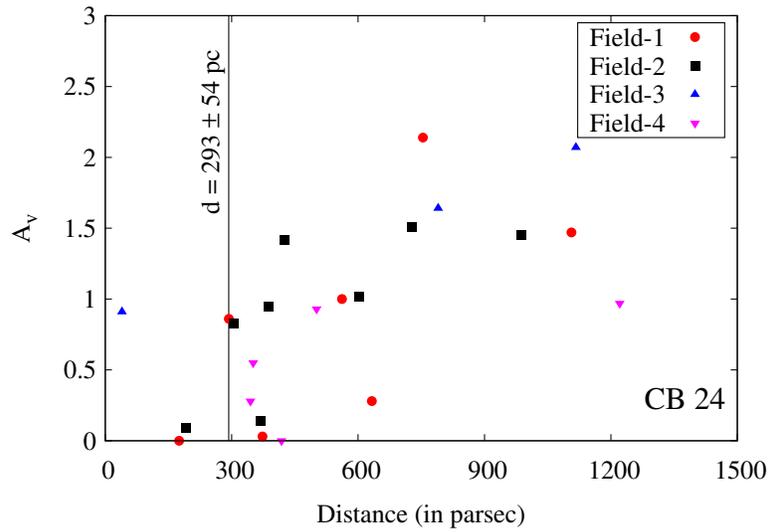}
 \caption{Extinction ($A_V$) versus distance for stars for four different fields in the vicinity of the CB 24. The vertical solid line is drawn at a distance of 293 parsec (star \# 10) where sudden rise in the $A_V$ occurs.}
 \end{center}
\end{figure}


\subsection{CB 188:}
CB 188 is a small, roundish and opaque globule which is located toward Aquila at a distance of $\approx 300$ pc and is associated with the Lindblad ring \citep{lh}. The dense core of CB 188  is associated with a cold \textit{IRAS} point source (19179+1129).

The field stars in CB 188 cloud are shown in Fig. 6. The extinction and distance obtained from this study are shown in Table-4. The distance versus extinction plot is shown in Fig. 7. It can be seen that star \# 5 shows sudden rise of extinction at $ d = 262 \pm 49$ pc. This is close to the distance suggested by \cite{lh}.

\begin{table}
\begin{center}
\caption{Selected field stars in CB 188.}
\vspace{1cm}
\begin{tabular}{|c|c|c|c|c|}
 \hline
	   \# & 2MASS &  $A_v$ & $d$ & Field \\	
	      &       & (mag) & (in pc)& \\	
	 \hline	
1	&	19195676+1131381	&	2.34	&	676	$\pm$	125	&	F4	\\
2	&	19200571+1131464	&	1.13	&	868	$\pm$	160	&	F4	\\
3	&	19201090+1133118	&	2.67	&	731	$\pm$	135	&	F4	\\
4	&	19201277+1139017	&	1.63	&	621	$\pm$	115	&	F1	\\
5	&	19201490+1140375	&	1.70    &	262	$\pm$	49	&	F1	\\
6	&	19201838+1133260	&	0	    &	492	$\pm$	91	&	F3	\\
7	&	19201988+1134172	&	2.60    &	877	$\pm$	162	&	F3	\\
8	&	19202266+1134140	&	1.97	&	486	$\pm$	90	&	F3	\\
9	&	19202307+1138353	&	0	    &	326	$\pm$	60	&	F2	\\
10	&	19202338+1135144	&	1.29	&	352	$\pm$	65	&	F3	\\
11	&	19202569+1140587	&	0.91	&	515	$\pm$	95	&	F2	\\
12	&	19203042+1140105	&	1.85	&	661	$\pm$	122	&	F2	\\
  \hline

\end{tabular}
\end{center}
\end{table}

\begin{figure}
  \hspace*{-2cm}
\includegraphics[width=220mm]{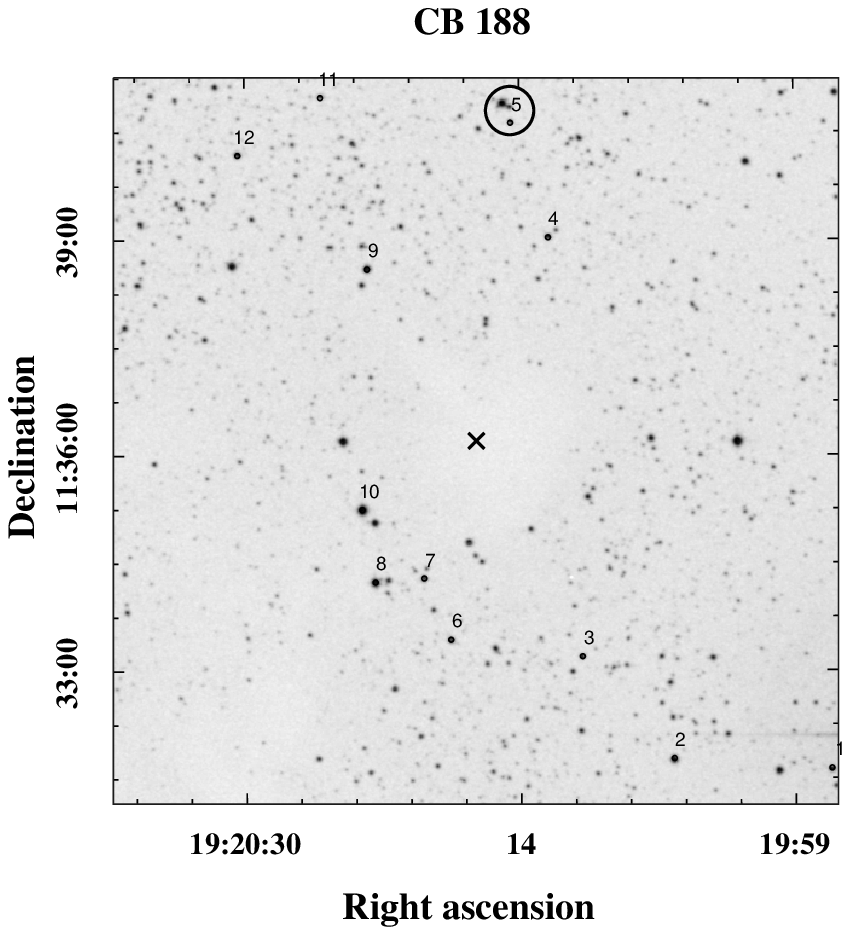}
 \caption{Selected 12 field stars in the vicinity of CB 188 shown on a $10 \times 10$ arcmin$^2$ R-band DSS image of the field. The `$\times$' symbol denotes the central coordinates of the cloud. Star \# 5 is marked by a circle which is the distance indicator of the cloud that shows sudden rise in the extinction.}
\end{figure}

\begin{figure}
\begin{center}
\includegraphics[width=100mm]{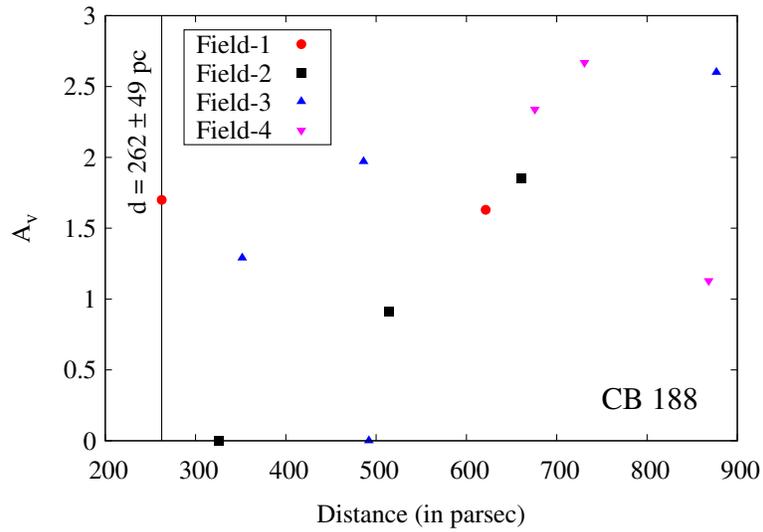}
 \caption{Extinction ($A_V$) versus distance for stars for four different fields in the vicinity of the CB 188. The vertical solid line is drawn at a distance of 262 parsec (star \# 5) where sudden rise in the $A_V$ occurs.}
 \end{center}
\end{figure}


\subsection{CB 224:}
CB 224 is a small, roundish globule located in the northern Cygnus region. The distance of this cloud was suggested by \cite{ln} to be $\approx 400$ parsec from the work of \cite{l1}. The core of this cloud contains two millimeter sources.  The field stars in the vicinity of the cloud CB 224 are shown in Fig. 8. The $A_V$ versus $d$ plot is shown in Fig. 9 which is based on results obtained from this work displayed in Table-5. We notice from Fig. 9 that the  star \# 8 ($A_V < 0.5$) with a distance of 228 pc could not be a distance indicator as three zero extincted stars could be seen in the Field-3 between 228 pc and 280 pc. It is also seen  that sudden rise in the extinction is noticed for star \#1 at $d = 378$ pc which is the distance to the cloud. This distance is comparable with the distance suggested by \cite{ln} which is 400 pc.

\begin{table}
\begin{center}
\caption{Selected field stars in CB 224.}
\vspace{1cm}
\begin{tabular}{|c|c|c|c|c|c|}
 \hline
	   \# & 2MASS &  $A_v$ & $d$ & Field \\	
	      &       & (mag) & (in pc)& \\	
	 \hline	
1	&	20353557+6349172	&	0.67	&	378	$\pm$	70	&	F4	 \\
2	&	20353770+6349063	&	0.06	&	367	$\pm$	68	&	F4	 \\
3	&	20360386+6358078	&	0.77	&	540	$\pm$	100	&	F1	 \\
4	&	20361149+6350051	&	1.66	&	829	$\pm$	153	&	F4	 \\
5	&	20361348+6349489	&	0.21	&	494	$\pm$	91	&	F4	 \\
6	&	20362586+6352534	&	0.39	&	418	$\pm$	77	&	F3	 \\
7	&	20362710+6352072	&	0    	&	274	$\pm$	51	&	F3	 \\
8	&	20362988+6357000	&	0.28	&	228	$\pm$	42	&	F2	 \\
9	&	20364079+6357001	&	0	    &	426	$\pm$	79	&	F2	 \\
10	&	20364079+6351004	&	0.09	&	335	$\pm$	62	&	F3	 \\
11	&	20364514+6350026	&	0	    &	266	$\pm$	49	&	F3	 \\
12	&	20364898+6353145	&	0.64	&	391	$\pm$	72	&	F2	 \\
13	&	20364923+6350288	&	0.38	&	427	$\pm$	79	&	F3	 \\
14	&	20365206+6349283	&	0	    &	257	$\pm$	47	&	F3	 \\
15	&	20365511+6351567	&	0	    &	418	$\pm$	77	&	F3	 \\
  \hline

\end{tabular}
\end{center}
\end{table}

\begin{figure}
  \hspace*{-2cm}
\includegraphics[width=220mm]{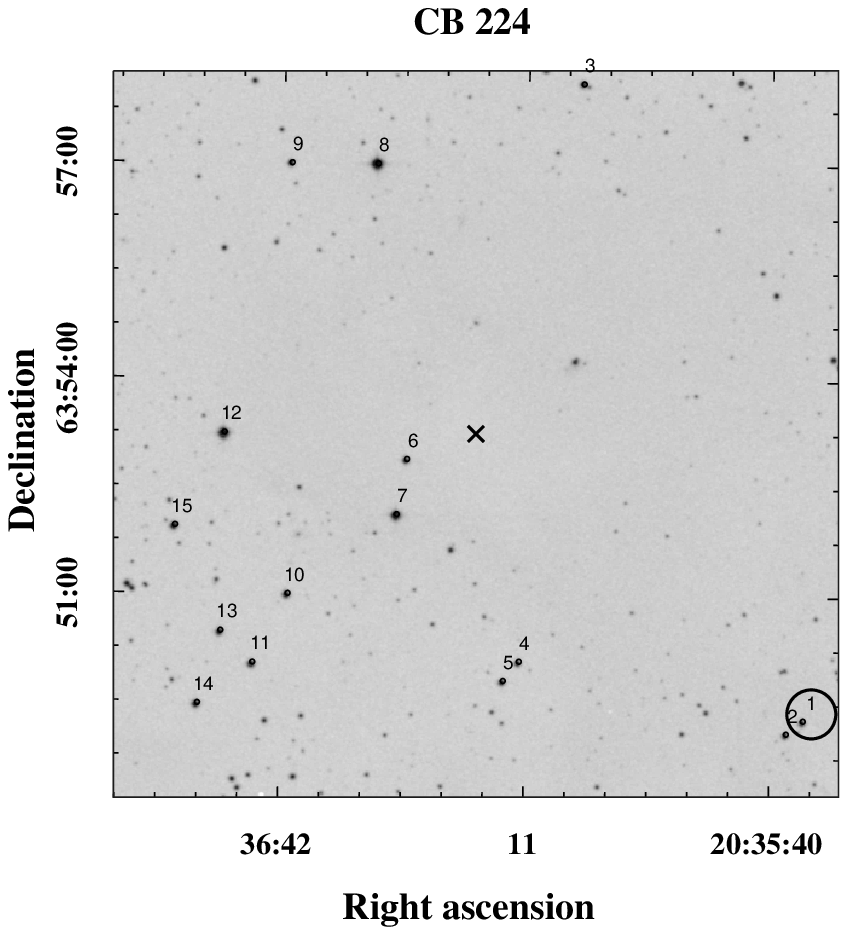}
 \caption{Selected 15 field stars in the vicinity of CB 224 shown on a $10 \times 10$ arcmin$^2$ R-band DSS image of the field. The `$\times$' symbol denotes the central coordinates of the cloud. Star \# 1 is marked by a circle which is the distance indicator of the cloud that shows sudden rise in the extinction.}
\end{figure}

\begin{figure}
\begin{center}
\includegraphics[width=100mm]{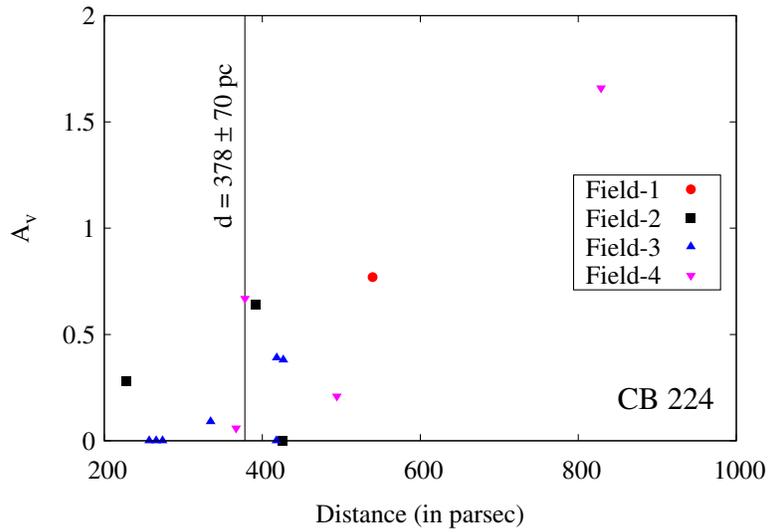}
 \caption{Extinction ($A_V$) versus distance for stars for four different fields in the vicinity of the CB 224. The vertical solid line is drawn at a distance of 378 parsec (star \# 1) where sudden rise in the $A_V$ occurs.}
 \end{center}
\end{figure}


\subsection{CB 230:}
CB 230 is a small, slightly cometary shaped globule located in the Cepheus Flare region whose distance was suggested by \cite{ln} to be  $\approx 400 \pm 100$ parsec from the result of \cite{wlh}. This cloud is connected with a reflection nebula VDB 141. A small aggregate can be found close to CB 230 and the member of this groups are [K98c] Em* 53, [K98c] Em* 58, 2MASS 21225427+6921345, the HAe star BD +68$^\circ$1118, and a candidate pre-main-sequence (PMS) star, 2MASS 21223461+6921142 (Kun et al. 2009 and references therein). \cite{kzs} determined the distance to this aggregate to be 390 pc by the assumption that BD +68$^\circ$1118 is a member and lies on the zero-age main sequence (ZAMS). They also noticed that star  BD +67$^\circ$1300 (VDB 141) closed to CB 230 may provide another distance estimate. The distance to this star is estimated to be $\approx$ 370 pc.

This cloud contains a dense core that is associated with two NIR reflection nebulae. The field stars in the vicinity of CB 230 are shown in Fig. 10. The extinction and distance obtained from this work are shown in Table-6 and plotted in Fig. 11. There is no stars found in Field-3. It can be seen from figure that star \#3 shows sudden rise of extinction which is situated at a distance of $293 \pm 54$ pc.  The distance determined from this study differs from the result suggested  by \cite{ln} and \cite{kzs}. Actually the line of sight to CB 230 may contain more than one cloud layers \citep{kzs}, the obtained result is a lower limit for the distance of this globule. It is also important to mention that the distance analysis for this cloud is restricted to only ten field stars.

\begin{table}
\begin{center}
\caption{Selected field stars in CB 230.}
\vspace{1cm}
\begin{tabular}{|c|c|c|c|c|c|}
 \hline
	   \# & 2MASS &  $A_v$ & $d$ & Field \\	
	      &       & (mag) & (in pc)& \\	
	 \hline	
1	&	21165611+6817080	&	0.35	&	261	$\pm$ 	48	&	F4	 \\
2	&	21170142+6820520	&	0.71	&	841	$\pm$ 	155	&	F1	 \\
3	&	21170258+6821556	&	1.09	&	293	$\pm$ 	54	&	F1	 \\
4	&	21170655+6813519	&	0.21	&	451	$\pm$ 	83	&	F4	 \\
5	&	21171001+6822241	&	0.03	&	497	$\pm$ 	92	&	F1	 \\
6	&	21171457+6815444	&	0.73	&	463	$\pm$ 	85	&	F4	 \\
7	&	21172702+6821111	&	1.08	&	322	$\pm$ 	59	&	F1	 \\
8	&	21173480+6822458	&	1.00    &	382	$\pm$ 	70	&	F1	 \\
9	&	21174268+6821557	&	1.91	&	661	$\pm$ 	122	&	F1	 \\
10	&	21175805+6821428	&	0.97	&	368	$\pm$ 	68	&	F2	 \\
  \hline

\end{tabular}
\end{center}
\end{table}

\begin{figure}
  \hspace*{-2.5cm}
\includegraphics[width=220mm]{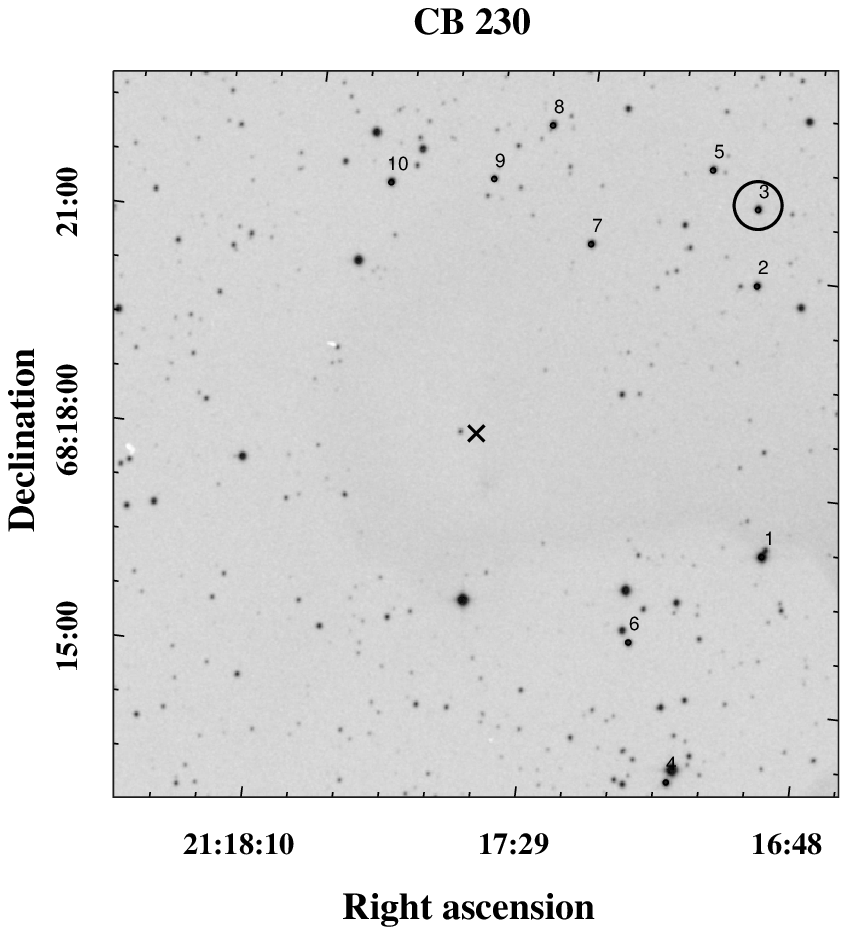}
 \caption{Selected 10 field stars in the vicinity of CB 230 shown on a $10 \times 10$ arcmin$^2$ R-band DSS image of the field. The `$\times$' symbol denotes the central coordinates of the cloud. Star \# 3 is marked by a circle which is the distance indicator of the cloud that shows sudden rise in the extinction.}
\end{figure}

\begin{figure}
\begin{center}
\includegraphics[width=100mm]{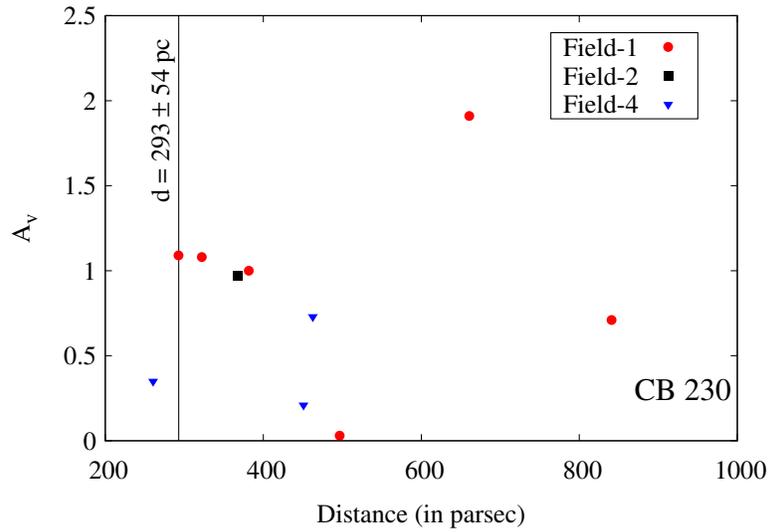}
 \caption{Extinction ($A_V$) versus distance for stars for three different fields in the vicinity of the CB 230. The vertical solid line is drawn at a distance of 293 parsec (star \# 3) where sudden rise in the $A_V$ occurs.}
 \end{center}
\end{figure}


\subsection{CB 240:}
This cloud is a small, roundish globule with a diffuse tail whose dense core is associated with an IRAS point source (22317+5816) \citep{hsw}. \cite{hsw} did not find any evidence for compact and embedded sources, but they noticed faint, extended emission from the surrounding dark cloud core. Since CB 240 is not associated with any known molecular cloud structure, \cite{lh} assigned the average distance of this cloud to be 500 pc, which is rather uncertain. We did not find the real distance of this cloud from literature survey. Thus the distance determination of CB 240 using NIR photometry will be the first such attempt.

The 55 field stars in CB 240 cloud is shown in Fig. 12. The extinction and distance determined for field stars using Maheswar's technique is shown in Table-7. We also plotted distance versus extinction plot for all the field stars and is shown in Fig. 13. The star \# 28 shows a sudden rise in extinction at a distance of $429 \pm 79$ pc which is the distance indicator of the cloud. The distance derived in this paper is the first real distance determination of CB 240.

\begin{table}
\begin{center}
\caption{Selected field stars in CB 240.}
\vspace{1cm}
\begin{tabular}{|c|c|c|c|c|c|}
 \hline
	   \# & 2MASS &  $A_v$ & $d$ & Field \\	
	      &       & (mag) & (in pc)& \\	
	 \hline	
1	&	22331564+5830286	&	0.59	&	519	$\pm$	96	&	F4	 \\
2	&	22331697+5833065	&	1.48	&	758	$\pm$	140	&	F4	 \\
3	&	22331821+5834513	&	3.11	&	986	$\pm$	182	&	F1	 \\
4	&	22331925+5831166	&	1.01	&	533	$\pm$	98	&	F4	 \\
5	&	22332007+5837427	&	2.43	&	888	$\pm$	164	&	F1	 \\
6	&	22332220+5829404	&	1.57	&	673	$\pm$	124	&	F4	 \\
7	&	22332437+5831123	&	1.63	&	378	$\pm$	70	&	F4	 \\
8	&	22332734+5829183	&	1.54	&	620	$\pm$	115	&	F4	 \\
9	&	22332734+5831369	&	1.05	&	853	$\pm$	158	&	F4	 \\
10	&	22333353+5829106	&	0.91	&	877	$\pm$	162	&	F4	 \\
11	&	22333445+5830305	&	1.30	&	358	$\pm$	66	&	F4	 \\
12	&	22333915+5830505	&	1.01	&	469	$\pm$	87	&	F4	 \\
13	&	22334055+5829188	&	0.20	&	505	$\pm$	93	&	F4	 \\
14	&	22334062+5830274	&	0.87	&	1166$\pm$	215	&	F4	 \\
15	&	22334104+5829369	&	1.86	&	943	$\pm$	174	&	F4	 \\
16	&	22334185+5831575	&	0	    &	286	$\pm$	53	&	F4	 \\
17	&	22334193+5830445	&	0.06	&	245	$\pm$	45	&	F4	 \\
18	&	22334416+5835039	&	2.67	&	492	$\pm$	91	&	F1	 \\
19	&	22334448+5829041	&	1.09	&	394	$\pm$	73	&	F4	 \\
20	&	22334522+5836428	&	1.95	&	1136$\pm$	210	&	F1	 \\
21	&	22334673+5830352	&	0.24	&	304	$\pm$	56	&	F4	 \\
22	&	22334849+5829212	&	0.90	&	1068$\pm$	197	&	F4	 \\
23	&	22335335+5837424	&	0	    &	219	$\pm$	40	&	F2	 \\
24	&	22335358+5830441	&	0.02	&	638	$\pm$	118	&	F3	 \\
25	&	22335364+5829372	&	0.81	&	511	$\pm$	94	&	F3	 \\
26	&	22335413+5835114	&	0	    &	191	$\pm$	35	&	F2	 \\
27	&	22335537+5836463	&	3.15	&	1358$\pm$	251	&	F2	 \\
28	&	22335729+5838067	&	2.67	&	429	$\pm$	79	&	F2	 \\
29	&	22335860+5833525	&	2.43	&	586	$\pm$	108	&	F2	 \\
30	&	22340075+5834226	&	1.41	&	586	$\pm$	108	&	F2	 \\
31	&	22340076+5830313	&	0.03	&	408	$\pm$	75	&	F3	 \\
32	&	22340508+5829327	&	1.58	&	1895$\pm$	350	&	F3	 \\
33	&	22340634+5831213	&	0	    &	495	$\pm$	91	&	F3	 \\
34	&	22340642+5828571	&	0.01	&	694	$\pm$	128	&	F3	 \\
35	&	22340778+5833312	&	0	    &	622	$\pm$	115	&	F3	 \\
36	&	22340958+5830451	&	1.90    &	1929$\pm$	356	&	F3	 \\
37	&	22340969+5835381	&	3.66	&	1996$\pm$	369	&	F2	 \\
38	&	22340969+5834154	&	1.56	&	1414$\pm$	262	&	F2	 \\
39	&	22341110+5833226	&	0.44	&	932	$\pm$	172	&	F3	 \\
40	&	22341173+5832257	&	0.07	&	1068$\pm$	197	&	F3	 \\
41	&	22341242+5831021 	&	0.55	&	973	$\pm$	180	&	F3	 \\
42	&	22341327+5829476	&	0.64	&	944	$\pm$	174	&	F3	 \\
43	&	22341422+5831272	&	1.15	&	763	$\pm$	141	&	F3	 \\
44	&	22341656+5834282	&	1.38	&	1155$\pm$	214	&	F2	 \\
45	&	22341782+5838256	&	0.79	&	474	$\pm$	88	&	F2	 \\
46	&	22341821+5835187	&	0.83	&	547	$\pm$	101	&	F2	 \\
47	&	22341825+5830369	&	0.79	&	648	$\pm$	120	&	F3	 \\
48	&	22341943+5831270	&	0.56	&	464	$\pm$	86	&	F3	 \\
49	&	22342030+5833565	&	0.04	&	850	$\pm$	157	&	F2	 \\
50	&	22342095+5837382	&	1.75	&	879	$\pm$	162	&	F2	 \\
51	&	22342135+5832537	&	0.57	&	885	$\pm$	164	&	F3	 \\
52	&	22342345+5832520	&	0.49	&	422	$\pm$	78	&	F3	 \\
53	&	22342408+5829566	&	0	    &	529	$\pm$	98	&	F3	 \\
  \hline

\end{tabular}
\end{center}
\end{table}

\begin{figure}
  \hspace*{-1cm}
\includegraphics[width=220mm]{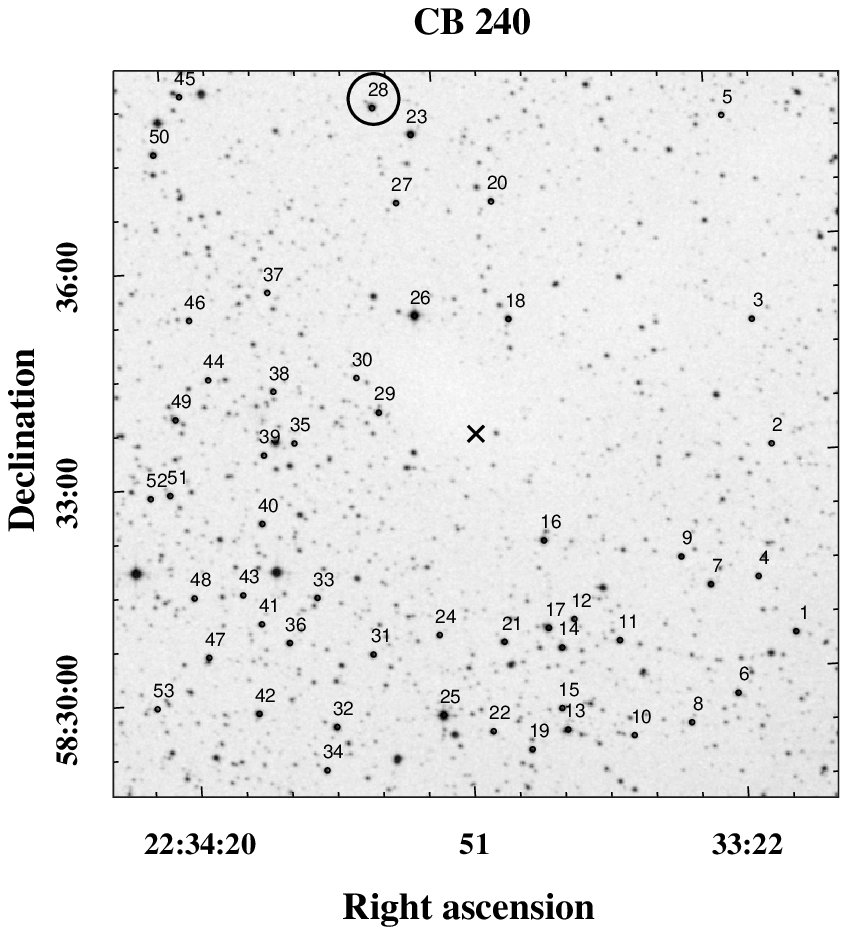}
 \caption{Selected 55 field stars in the vicinity of CB 240 shown on a $10 \times 10$ arcmin$^2$ R-band DSS image of the field. The `$\times$' symbol denotes the central coordinates of the cloud. Star \# 28 is marked by a circle which is the distance indicator of the cloud that shows sudden rise in the extinction.}
\end{figure}


\begin{figure}
\begin{center}
\includegraphics[width=100mm]{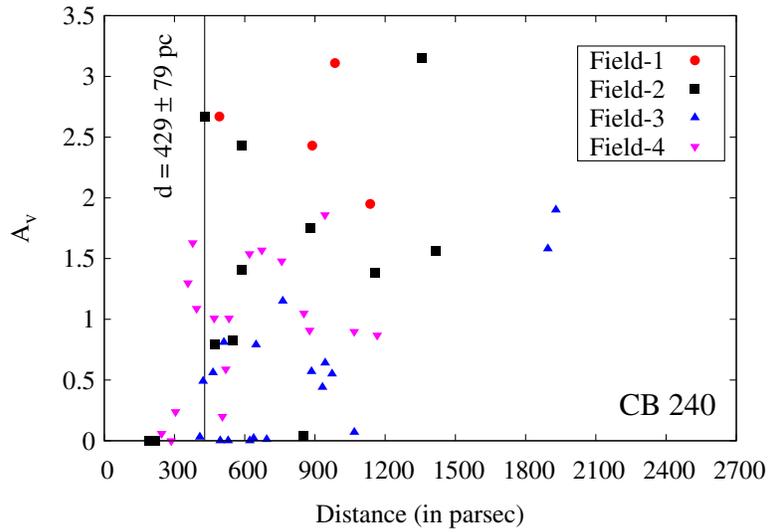}
 \caption{Extinction ($A_V$) versus distance for stars for four different fields in the vicinity of the CB 240 cloud. The vertical solid line is drawn at a distance of 429 parsec (star \# 28) where sudden rise in the $A_V$ occurs.}
 \end{center}
\end{figure}

\section{Summary}
 The distances to six selected clouds CB 17, CB 24, CB 188, CB 224, CB 230 and CB 240 are estimated to be $478 \pm 88$, $293 \pm 54$, $262 \pm 49$, $378 \pm 70$, $293 \pm 54$ and $429 \pm 79$ parsec respectively. It is to be noted that the distances of clouds CB 24 and CB 230 are same. The typical error in determination of distances to clouds are found to be $\sim 18 \%$. The distances obtained from this study and reported by other investigators are compiled in Table-8.  The distances to four clouds CB 17, CB 188, CB 224 and CB 230 reported in literature were indirectly derived with a method which associates the clouds with larger molecular clouds or stars. However, direct estimation of distance to CB 24 was made by \cite{pc}  by identifying M dwarfs lying both in front of and behind the cloud. CB 240 is not associated with any known molecular cloud structure, the average distance of this cloud was assumed to be 500 pc by \cite{lh}, which is rather uncertain. We found that the distances of clouds CB 24, CB 188 and CB 224 are almost comparable with literature. Since the line of sight to CB 230 may contain more than one cloud layers, the result obtained from this analysis is a lower limit for the distance of this globule. Further, the distance estimated for CB 240 in the present work is the first real distance determination of this cloud. It has been already shown by Maheswar et al. (2010, 2011) that this technique revealed the distances accurately for several globules, because the distance obtained from this study is related with field stars in the vicinity of the cloud. But it is also important to mention that the present analysis is restricted to a limited number of data points (especially for clouds CB 188, CB 224 and CB 230).

\begin{table}
\begin{center}
\caption{Distances of the selected clouds compiled from the literature along with the distances obtained from the present work.}
\vspace{1cm}
\begin{tabular}{|c|l|c|l|}
  \hline
  S/N & Cloud &  $d$  (literature) & $d$ (this work)  \\
      &       &   (in pc) & (in pc)\\
  \hline
  1 & CB 17   & $250 \pm 50$ & $478 \pm 88$ \\
  2 & CB 24   & 360$^*$ & $293 \pm 54$ \\
  3  & CB 188 & 300 & $262 \pm 49$\\
  4 & CB 224  & 400  & $378 \pm 70$ \\
  5 & CB 230   & 400 & $293 \pm 54$ \\
  6 & CB 240   & 500$^\dag$ & $429 \pm 79$ \\
  \hline
\end{tabular}
\end{center}
\begin{center}
* Maximum distance

$^\dag$ Uncertain
\end{center}
\end{table}

\section*{ACKNOWLEDGMENTS}

This work makes use of data products from the Two Micron All Sky Survey (2MASS),
which is a joint project of the University of Massachusetts and
the Infrared Processing and Analysis Center/California Institute
of Technology, funded by the National Aeronautics and Space
Administration and the National Science Foundation. The reviewer of this paper is highly
acknowledged for his constructive comments and suggestions which definitely help to improve the quality of the
paper. This work is supported by the Science and Engineering Research Board (SERB), a statutory body under Department
of Science and Technology (DST), Government of India, under Fast Track scheme for
Young Scientist (SR/FTP/PS-092/2011).

\label{lastpage}

\end{document}